\documentclass[prl,twocolumn,groupedaddress]{revtex4-1}
\catcode`\@=11
\def\lsim{\mathrel{\mathpalette\@versim<}}
\def\gsim{\mathrel{\mathpalette\@versim>}}
\def\@versim#1#2{\vcenter{\offinterlineskip
\ialign{$\m@th#1\hfil##\hfil$\crcr#2\crcr\sim\crcr } }}
\catcode`\@=12

\catcode`@=12
\usepackage[export]{adjustbox}	
\usepackage{float}
\usepackage{caption}
\usepackage{subcaption}
\usepackage{epsfig,bbm}
\usepackage{amsmath}
\usepackage{slashed}
\usepackage{tikz}
\usepackage{bm}
\usepackage{amssymb}
\usepackage{mathtools}
\usepackage{graphicx}
\newcommand{\be}{\begin{equation}}
\newcommand{\ee}{\end{equation}}
\newcommand{\bea}{\begin{eqnarray}}
\newcommand{\eea}{\end{eqnarray}}

\begin{document}
\thispagestyle{empty}
\begin{flushright}
\end{flushright}

\begin{center}
\title{The pole inflation from broken non-compact isometry in Weyl gravity
}
\author{Hyun Min Lee}
\email[]{hminlee@cau.ac.kr}
\affiliation{Department of Physics, Chung-Ang University, Seoul
06974, Korea}
\begin{abstract}
We propose the microscopic origin of the pole inflation from the scalar fields of broken non-compact isometry in Weyl gravity.  We show that the $SO(1,N)$ isometry in the field space in combination with the Weyl symmetry relates the form of the non-minimal couplings to the one of the potential in the Jordan frame.  In the presence of an explicit breaking of the $SO(1,N)$ symmetry in the coefficient of the potential, we realize the pole inflation near the pole of the inflaton kinetic term. Applying our results to the Higgs or PQ inflation models, we find that there is one parameter family of the solutions for the pole inflation, depending on the overall coefficient of the Weyl covariant derivatives for scalar fields. The same coefficient not only makes the predictions of the pole inflation varying, being compatible with the Planck data, but also determines the mass of the Weyl gauge field.  We also show that the isocurvature perturbations of the axion can be suppressed sufficiently during the PQ pole inflation, and the massive Weyl gauge field produced during reheating serves as a dark matter candidate.
\end{abstract}
\maketitle
\end{center}

\section{Introduction}
%

Inflation has been a main paradigm for the early universe by which various problems in the Standard Big Bang cosmology are solved and the initial conditions for the flat, homogeneous and isotropic universe can be explained. A slowly rolling scalar field, the so called inflaton, is required to derive inflation and its quantum fluctuations generate necessary inhomogeneities observed in the Cosmic Microwave Background (CMB) and the large scale structures. The Higgs inflation with a non-minimal coupling \cite{Higgsinflation} has drawn a lot of attention because it shows a possibility that inflation is realized by the Higgs field within the Standard Model (SM), but a consistent picture beyond the Higgs inflation should emerge due to the problem with a large non-minimal coupling \cite{unitarity,sigmamodels,conformal}.

There is another class of inflation models where  the inflaton has a conformal coupling to gravity \cite{linde,conformal}, so that inflation takes place close to the pole of the inflaton kinetic term in the Einstein frame This is dubbed the pole inflation. The concept of the pole inflation for $\alpha$-attractor models was also introduced in Ref.~\cite{Kallosh:2013yoa,Galante:2014ifa,Broy:2015qna,Terada:2016nqg}. The global conformal symmetry can be gauged by a local conformal symmetry or Weyl symmetry. As a result, the Planck mass can be generated dynamically from the vacuum expectation value of the dilaton field or one of the scalar fields of a non-linear sigma model type \cite{Weyl0}. Furthermore, the multi-field models with Weyl symmetry including the SM Higgs were considered for inflation \cite{Weyl1,Weyl2,Weyl3}.

The SM is based on the gauge principle explaining the forces in nature after the gauge symmetries are broken spontaneously, and there is an approximate custodial symmetry for the SM Higgs fields, which is broken only by the $U(1)_Y$ gauge coupling and the Yukawa couplings. A similar gauge principle is applied to the theory of gravitation such that the conformal symmetry is gauged and it is broken spontaneously. As a result, the Einstein gravity is reproduced, up to a massive Weyl gauge field which couples to gravity minimally.  The goal of this article is to make the Weyl gauge symmetry manifest in the extension with extra scalar multiplets beyond the dilaton and the metric tensor, so it is suitable for a unified description of the gravity-Higgs system based on both the gauge symmetry principle and the extended custodial symmetry for the dilaton and the extra multiple fields. The scalar sector contains an extended custodial symmetry for the SM Higgs or the Peccei-Quinn (PQ) fields, which is the inflaton candidate with an appropriate form of the scalar potential in the context of the pole inflation. The full content of the SM or its non-gravitational extensions can be easily accommodated in this setup. 

In this article, we propose the multi-field models for inflation respecting both the Weyl symmetry and the broken non-compact isometry in the field space such as $SO(1,N)$, which is the extension of the isometry or custodial symmetry of the non-dilaton scalar fields. After the Weyl symmetry is broken spontaneously due to the VEV of the dilaton, the $SO(1,N)$ symmetry is spontaneously broken to $SO(N)$ and the Planck scale is generated. In this scenario, we pursue a concrete realization of the pole inflation in Weyl gravity, which is applicable to the cases with the SM Higgs fields \cite{Higgspole} and the PQ singlet scalar field \cite{Lee:2023dtw,Lee:2024bij} transforming under the electroweak symmetry or a global $U(1)$ PQ symmetry, respectively. For inflation, we introduce an explicit breaking for the $SO(1,N)$ symmetry only in the effective quartic coupling but the $SO(N)$ symmetry remains unbroken in the Lagrangian. We discuss the roles of the Weyl covariant derivatives for scalar fields for the mass of the Weyl gauge field and the solutions for the pole inflation.

\section{The setup}

We consider the dilaton $\chi$, an $N$-dimensional scalar multiplet, $\Phi=\frac{1}{\sqrt{2}}(\phi_1,\phi_2,\cdots, \phi_N)^T$, composed of $N$ real scalar fields, and the Weyl gauge field $w_\mu$ in Weyl gravity. It can accommodate the SM Higgs doublet for $N=4$ or the PQ singlet scalar field for $N=2$.
Then, the Jordan frame Lagrangian for bosons respecting the Weyl invariance and the $SO(1,N)$ isometry, in $\{\chi, \phi_i\}$ with $i=1,2,\cdots, N$, is given by
\bea
\frac{{\cal L}_J}{\sqrt{-g_J}}&=&(1+a)\bigg[-\frac{1}{12}  (\chi^2-\phi_i^2 ) R-\frac{1}{2}(\partial_\mu\chi)^2+\frac{1}{2}(\partial_\mu \phi_i)^2 \bigg]  \nonumber \\
&&+ \frac{1}{2}a(D_\mu\chi)^2 -\frac{1}{2}a (D_\mu\phi_i)^2 -\frac{1}{4} w_{\mu\nu}w^{\mu\nu} - V\label{Lag}
\eea 
with
\bea
V(\chi, \phi_i)= \frac{1}{\langle\chi^4\rangle} f(\phi^2_i/\chi^2) (\chi^2-\phi_i^2 )^2.
\eea
Here, we note that the Weyl gauge transformations are 
\bea
&& g_{\mu\nu}\to e^{2\alpha(x)} g_{\mu\nu}, \, \chi\to e^{-\alpha(x)} \chi, \nonumber \\
&& \,\phi_i\to e^{-\alpha(x)} \phi_i,   \, w_\mu\to w_\mu-\frac{1}{g_w} \partial_\mu\alpha(x),
\eea
with $\alpha(x)$ being an arbitrary real transformation parameter. 
Then, the covariant derivatives for the dilaton and the Higgs fields are given by 
\bea
D_\mu\chi=(\partial_\mu - g_w w_\mu )\chi, \quad D_\mu \phi_i=(\partial_\mu - g_w w_\mu )\phi_i,
\eea
with $g_w$ being the Weyl gauge coupling,
and the field strength tensor for the Weyl gauge field is $w_{\mu\nu}=\partial_\mu w_\nu-\partial_\nu w_\mu$.  We normalized the scalar kinetic terms in eq.~(\ref{Lag}), up to a constant parameter $a$. We didn't include the SM gauge interactions for the Higgs fields explicitly, but they can be also introduced easily. 
 $ f(\phi^2_i/\chi^2)$ is an arbitrary function of $\phi^2_i/\chi^2$, respecting the Weyl gauge symmetry, but it breaks the $SO(1,N)$ isometry down to $SO(N)$ explicitly. 
If $ f(\phi^2_i/\chi^2)$ is a constant parameter, the full $SO(1,N)$ is respected, but it leads to a constant vacuum energy after a gauge fixing, as will be shown later. Thus, in order to consider the inflationary cosmology with a slow-roll inflaton, $ f(\phi^2_i/\chi^2)$ must not be constant.

Due to the Weyl symmetry in the Jordan frame Lagrangin in eq.~(\ref{Lag}), the form of the Lagrangian is the same in any other frames related by the Weyl transformations unless a gauge for the Weyl symmetry is fixed. 
Thus, we first fix the gauge for the Weyl symmetry with $\chi=\langle \chi\rangle=\sqrt{6/(1+a)}$ in the Jordan frame, so we  break the Weyl symmetry and the $SO(1,N)$ symmetry spontaneously. 
Then, the Lagrangian in eq.~(\ref{Lag}) becomes
\begin{widetext}
\bea
\frac{{\cal L}_J}{\sqrt{-g_J}}&=&-\frac{1}{2}\bigg(1- \frac{1}{6}(1+a)\phi_i^2 \bigg) R+\frac{1}{2}(\partial_\mu \phi_i)^2+\frac{1}{2}a g_w w_\mu  \partial^\mu\phi^2_i   -\frac{1}{2}a g^2_w  \phi^2_i w_\mu w^\mu \nonumber \\
&&\quad -\frac{1}{4} w_{\mu\nu}w^{\mu\nu}+\frac{1}{2} m^2_w w_\mu w^\mu - V(\langle\chi\rangle, \phi_i) \label{gfLag}
\eea
\end{widetext}
with
\bea
m^2_w &=&a g^2_w \langle\chi^2\rangle=\frac{6ag^2_w}{1+a}, \label{wmass} \\
V(\langle\chi\rangle, \phi_i) &=& f(\phi^2_i/\langle\chi^2\rangle) \bigg(1- \frac{1}{6}(1+a)\phi_i^2 \bigg)^2.
\eea
In the presence of electroweak symmetry breaking, there is an additional contribution to the Weyl gauge field by $a g^2_w v^2$, but it is negligble as compared to the one from the dilaton VEV.

For $a=0$, we get the same form of the Higgs part of the Lagrangian as in the Higgs pole inflation where the Higgs is conformally coupled to gravity and both the effective Planck scale and the Higgs potential depend on the same factor, $\big(1- \frac{1}{6}\phi_i^2 \big)$ \cite{Higgspole}. But, in this case, the Weyl gauge field would be massless, while being decoupled from the Higgs fields.
However, for $a\neq 0$, the Weyl gauge field becomes massive and we can generalize the Higgs pole inflation, as will be discussed later.  As compared to the case with a conformal symmetry in Ref.~\cite{linde}, our results rely on the spontaneously broken Weyl gauge symmetry. Thus, there are extra interactions terms between the Weyl gauge field and the Higgs fields. The same results hold for the PQ pole inflation.

\section{Gauge-fixed Lagrangian in Einstein frame}

Using eq.~(\ref{gfLag}) and the redefined Weyl gauge field, we can rewrite the Lagrangian in the Jordan frame, as follows,
\bea
\frac{{\cal L}_{J,{\rm eff}}}{\sqrt{-g_J}}&=&-\frac{1}{2}\bigg(1- \frac{1}{6}(1+a)\phi_i^2 \bigg) R+\frac{1}{2}(\partial_\mu \phi_i)^2\nonumber \\
&&-\frac{1}{48} a(1+a)\,\cdot \frac{ (\partial_\mu\phi^2_i)^2}{1-\frac{1}{6}(1+a) \phi^2_i} \nonumber \\
&& -  f(\phi^2_i/\langle\chi^2\rangle) \bigg(1- \frac{1}{6}(1+a)\phi_i^2 \bigg)^2 \label{Jeff}  \\
&&-\frac{1}{4} {\widetilde w}_{\mu\nu}{\widetilde w}^{\mu\nu} +\frac{1}{2}m^2_w\bigg(1-\frac{1}{6}(1+a) \phi^2_i\bigg)   {\widetilde w}_\mu  {\widetilde w}^\mu \nonumber 
\eea
where 
\bea
  {\widetilde w}_\mu \equiv  w_\mu-\frac{1}{2g_w} \partial_\mu \ln   (m^2_w - a g^2_w \phi_i^2).
\eea
Here, we used eq.~(\ref{wmass}) for $m^2_w$. 

Now making a rescaling of the metric by $g_{\mu\nu,J}=g_{\mu\nu,E}/\Omega$ with $\Omega= 1- \frac{1}{6}(1+a)\phi_i^2$,
we obtain the Einstein frame Lagrangian from eq.~(\ref{Jeff}) as
\bea
\frac{{\cal L}_{E}}{\sqrt{-g_E}}&=& -\frac{1}{2} R + \frac{3}{4\Omega^2} \Big(\partial_\mu \Omega \Big)^2 + \frac{1}{2} \frac{(\partial_\mu\phi_i)^2}{\Omega} \nonumber \\ 
&&-\frac{1}{48} a(1+a)\,\cdot \frac{ (\partial_\mu\phi^2_i)^2}{\Omega^2} -  f(\phi^2_i/\langle\chi^2\rangle) \nonumber \\
&& -\frac{1}{4} {\widetilde w}_{\mu\nu}{\widetilde w}^{\mu\nu} +\frac{1}{2}m^2_w  {\widetilde w}_\mu  {\widetilde w}^\mu. 
\eea
Thus, the redefined Weyl gauge field ${\widetilde w}_\mu$ is decoupled from the Higgs fields and it couples to gravity minimally.
We note that the Weyl gauge field has an arbitrary mass depending the Weyl gauge coupling $g_w$ and $a$, and  there is a $Z_2$ symmetry for $ {\widetilde w}_{\mu}$ in the Lagrangian. So, the Weyl gauge field could be a good candidate for dark matter which is gravitationally produced during inflation or reheating.

From $\partial_\mu\Omega = -\frac{1}{6}(1+a)\partial_\mu \phi^2_i$, we can recast the Einstein frame Lagrangian without the Weyl gauge field in a simpler form,
\bea
\frac{{\cal L}_{E}}{\sqrt{-g_E}}&=& -\frac{1}{2} R + \frac{1}{2}\frac{(\partial_\mu\phi_i)^2}{\big(1-\frac{1}{6} (1+a) \phi^2_i \big)^2} \label{LagE} \\
&&+\frac{1}{12} (1+a)\cdot \frac{\frac{1}{4}(\partial_\mu\phi^2_i)^2-\phi^2_j (\partial_\mu\phi_i)^2}{\big(1-\frac{1}{6} (1+a) \phi^2_i \big)^2}-  V_E(\phi_i)  \nonumber 
\eea
with $V_E(\phi_i)=f(\phi^2_i/\langle\chi^2\rangle)$.
Therefore, the Higgs kinetic terms in the above Lagrangian are of the same form as in the Higgs pole inflation \cite{Higgspole}, except with an arbitrary parameter $a$. 

For the pole inflation, we take the coefficient of the Jordan frame potential as
\bea
f(\phi^2_i/\chi^2) =V_0+ \frac{1}{2}m^2_\phi \langle\chi^2\rangle\cdot \frac{\phi^2_i}{\chi^2}+ \frac{1}{4}\lambda_\phi \langle\chi^4\rangle\cdot \frac{(\phi^2_i)^2}{\chi^4}.
\eea
Here, $V_0$ corresponds to the vacuum energy, which respects the $SO(1,N)$ symmetry, but $m^2_\phi, \lambda_\phi$ terms break the $SO(1,N)$ symmetry into $SO(N)$. 
Then, under the gauge condition, $\chi=\langle \chi\rangle=\sqrt{6/(1+a)}$, it leads to the standard form of the Higgs-like potential in eq.~(\ref{LagE}) as
\bea
V_E(\phi_i)=\frac{1}{2}m^2_\phi \phi^2_i + \frac{1}{4}\lambda_\phi (\phi^2_i)^2. \label{Epot}
\eea

\section{Pole inflation models in Weyl gravity}

From the general Einstein frame Lagrangian obtained in the previous section, we discuss the generalization of the pole inflation scenarios with the SM Higgs doublet or the PQ singlet scalar field in Weyl gravity. 

{\bf\it Generalized Higgs pole inflation}:
We realize the pole inflation with the SM Higgs inflation \cite{Higgspole} as an example for $N=4$.
In unitary gauge, the Higgs fields composing an $SU(2)_L$ doublet, take $\phi_1=h$ and $\phi_2=\phi_3=\phi_4=0$, so the second kinetic term in eq.~(\ref{LagE}) vanishes.
Then, the Einstein frame Lagrangian in eq.~(\ref{LagE}) becomes
\bea
\frac{{\cal L}_{E}}{\sqrt{-g_E}}=-\frac{1}{2} R + \frac{1}{2}\frac{(\partial_\mu h)^2}{\big(1-\frac{1}{6} (1+a) h^2 \big)^2} - V_E(h) \label{Lagh}
\eea
with $V_E(h)= \frac{1}{2}m^2_H  h^2 + \frac{1}{4}\lambda_H h^2$ after $m^2_\phi, \lambda_\phi$ in eq.~(\ref{Epot}) being replaced by the Higgs parameters, $m^2_H, \lambda_H$, respectively. This takes precisely the same form as in the Higgs pole inflation \cite{Higgspole}, again except the parameter $a$. For $m^2_H<0$ and $\lambda_H>0$, the VEV of the Higgs is determined by $\langle h\rangle =v=\sqrt{-m^2_H/\lambda_H}$.

As a result, making the Higgs kinetic term canonical for
\bea
h= \langle\chi\rangle \tanh \bigg(\frac{\psi}{\langle\chi\rangle}\bigg), \label{redef}
\eea
we obtain the inflaton Lagrangian in eq.~(\ref{Lagh}) as
\bea
\frac{{\cal L}_{E}}{\sqrt{-g_E}}= -\frac{1}{2} R + \frac{1}{2} (\partial_\mu\psi)^2 -V_E(\psi) \label{Einact}
\eea
where the inflaton potential with the Higgs quartic coupling only becomes
\bea
V_E(\psi) =9 \lambda_H \tanh^4  \bigg(\frac{\psi}{\langle\chi\rangle}\bigg). \label{Vinf}
\eea
Then, the Higgs pole inflation corresponds to $a=0$ or $\langle\chi\rangle=\sqrt{6}$.
As compared to $T$ $\alpha$-attractor models with the potential $V_E\sim \tanh^{2n}(\psi/\sqrt{6\alpha})$ \cite{Kallosh:2013yoa}, we can identify the model parameters by $1+a=\frac{1}{\alpha}$ and $n=2$. Thus, for $a=[0,1]$, the $T$ $\alpha$-attractor models vary by $\alpha=[1,\frac{1}{2}]$.

We note that the interaction Lagrangian of the Higgs boson to the electroweak bosons in unitary gauge in Einstein frame contains
\bea
{\cal L}_{h,{\rm gauge}}=F(h) \,\Big(2g^2 W_\mu W^\mu + (g' B_\mu -g W^3_\mu)^2)\Big),
\eea
with 
\bea
F(h)&\equiv &\frac{h^2}{8(1-\frac{1}{6}(1+a)h^2)} \nonumber \\
&=&\frac{1}{8} \langle\chi\rangle^2 \sinh^2 \bigg(\frac{\psi}{\langle\chi\rangle}\bigg),
\eea
where we used eq.~(\ref{redef}) in the second line.
Then, during inflation with $\psi\gg \langle\chi\rangle$, the effective masses for the electroweak gauge bosons  are given by $M^2_W\gg \frac{1}{4}g^2M^2_P$ and $M^2_Z\gg \frac{1}{4}(g^2+g^{\prime 2})M^2_P$, so they are safely decoupled from the inflaton. On the other hand, after inflation, $\psi\ll \langle\chi\rangle$, for which $ \langle\chi\rangle^2 \tanh^2 \big(\frac{\psi}{\langle\chi\rangle}\big) \simeq \psi^2 $, so we can recover the standard interactions of the Higgs boson to the electroweak gauge bosons, so reheating can proceed. 

From the slow-roll parameters with eq.~(\ref{Einact}), we get the spectral index and the tensor-to-scalar ratio at horizon exit  in terms of the number of efoldings, as follows, 
\bea
n_s&=& 1-\frac{64(8N+\langle\chi^2\rangle)}{256N^2-\langle\chi^4\rangle}, \label{sindex} \\
r&=&  \frac{512 \langle\chi^2\rangle}{256N^2-\langle\chi^4\rangle}. \label{ratio}
\eea

\begin{figure}[t]
\centering
\includegraphics[width=0.45\textwidth,clip]{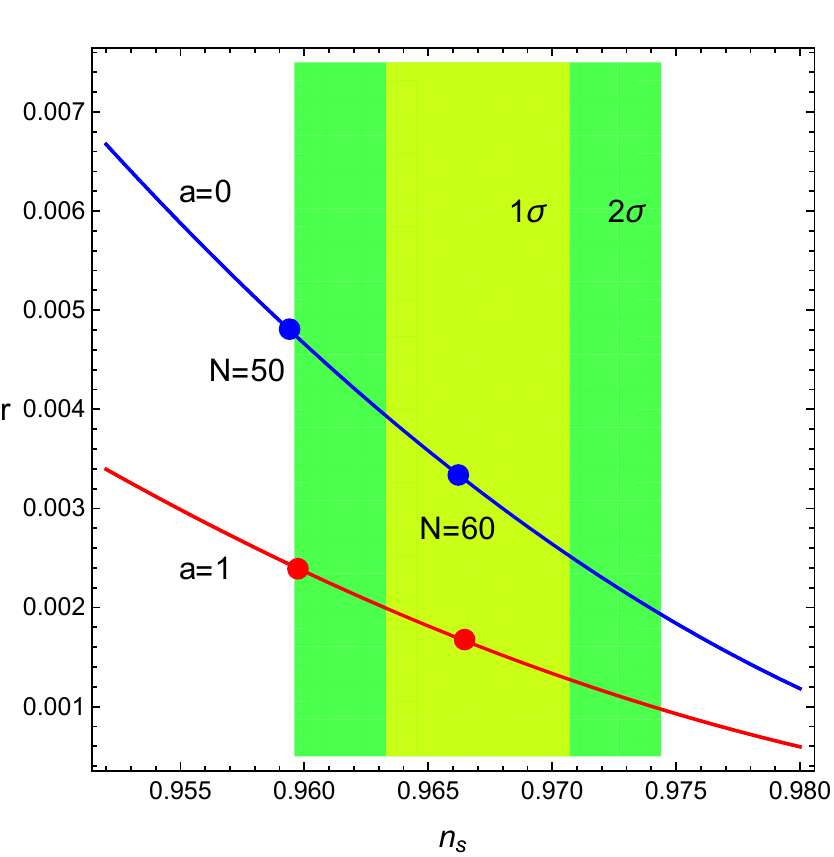}
\caption{Spectral index $n_s$ vs tensor-to-scalar ratio $r$ for the pole inflation in Weyl gravity. Blue and red solid lines are the cases wirh $a=0,1$, respectively, and the blue or red bullets indicate the boundaries where the number of efoldiings is given by $N=50, 60$. The Planck bounds on the spectral index within $1\sigma$ and $2\sigma$ errors are shown in yellow and green regions, respectively.
} 
\label{Fig:ns}
\end{figure}

As a result, from eq.~(\ref{sindex}), we obtain the spectral index as $n_s=0.9662-0.9666$ for $N=60$ and $a=[0,1]$, which agrees with the observed spectral index from Planck, $n_s=0.967\pm 0.0037 $ \cite{planck}.
Moreover, we also predict  the tensor-to-scalar ratio as $r=0.00083-0.0033$ for $N=60$ and $a=[0,1]$, which is compatible with  the bound from the combined Planck and Keck data \cite{keck}, $r<0.036$. We also find that the CMB normalization, $A_s=\frac{1}{24\pi^2} \frac{V_I}{\epsilon_*}=2.1\times 10^{-9}$,  sets the inflation energy scale by
\bea
\lambda_H = (3.4\times 10^{-9}) \,r. \label{CMB}
\eea
Thus, for a given $r$, we need the Higgs quartic coupling during inflation to be $\lambda_H=2.8\times 10^{-12}-1.1\times 10^{-11}$.
Such a tiny quartic coupling for the SM Higgs could be achieved when the corresponding beta function is sufficiently small in the presence of the couplings of singlet scalar fields to the SM Higgs \cite{Higgspole}.

In Fig.~\ref{Fig:ns}, we depict the inflationary predictions of the pole inflation in Weyl gravity in the spectral index $n_s$ vs the tensor-to-scalar ration $r$. We show the results for $a=0$ and $a=1$ in blue and red lines, respectively, while the number of efoldings is bounded between $N=50$ and $60$ at the pairs of blue or red bullets. We overlay the bounds from Planck on the spectral index  within $1\sigma$ and $2\sigma$ errors in yellow and green regions, respectively.

{\bf\it Generalized PQ pole inflation}: We now realize the pole inflation with the PQ singlet scalar field \cite{Lee:2023dtw,Lee:2024bij} as an example with a complex scalar field, that is, $N=2$.
In this case, a PQ complex scalar field, $\Phi=\frac{1}{\sqrt{2}}(\phi_1+i\phi_2)$, transforms under the global $U(1)$ PQ symmetry.
Taking the PQ field in the polar representation, $\Phi=\frac{1}{\sqrt{2}}\rho\, e^{i\theta}$, the Einstein frame Lagrangian in eq.~(\ref{LagE}) becomes
\bea
\frac{{\cal L}_{E}}{\sqrt{-g_E}}&=&-\frac{1}{2} R +\frac{1}{2} \frac{(\partial_\mu\rho)^2}{\big(1-\frac{1}{6} (1+a) \rho^2 \big)^2}\nonumber \\
&&+\frac{1}{2}(1+a)\cdot\,\frac{\rho^2(\partial_\mu\theta)^2}{\big(1-\frac{1}{6} (1+a) \rho^2 \big)} - V_E(\rho) \label{PQLagE}
\eea
where $V_E(\rho)=\frac{1}{2}m^2_\Phi \rho^2+\frac{1}{4}\lambda_\Phi \rho^4$ after $m^2_\phi, \lambda_\phi$ in eq.~(\ref{Epot}) being replaced by the PQ parameters, $m^2_\Phi, \lambda_\Phi$, respectively. This takes precisely the same form as in the PQ pole inflation \cite{Lee:2023dtw,Lee:2024bij}, again except for the parameter $a$. For $m^2_\Phi<0$ and $\lambda_\Phi>0$, the VEV of the PQ field is determined by $\langle\Phi\rangle =f_a=\sqrt{-m^2_\Phi/\lambda_\Phi}$.

Making the kinetic term for the radial mode canonically normalized by
\bea
\rho=\langle\chi\rangle \tanh \bigg(\frac{\psi}{\langle\chi\rangle}\bigg),
\eea
we obtain eq.~(\ref{PQLagE}) as
\bea
\frac{{\cal L}_{E}}{\sqrt{-g_E}}&=&-\frac{1}{2} R +\frac{1}{2}(\partial_\mu\psi)^2 \nonumber \\
&&+3 \sinh^2  \bigg(\frac{\psi}{\langle\chi\rangle}\bigg)\, (\partial_\mu\theta)^2 -V_E(\psi)
\eea
where the inflaton potential with the PQ quartic coupling only is given by
\bea
V_E(\psi)= 9 \lambda_\Phi \tanh^4  \bigg(\frac{\psi}{\langle\chi\rangle}\bigg).
\eea
Therefore, the same inflationary predictions as in the Higgs pole inflation are maintained, as far as the Higgs quartic coupling in eq.~(\ref{CMB}) is replaced by the PQ quartic coupling. In this case, a similarly tiny quartic coupling for the PQ field can be stable under the renormalization group running for small Yukawa couplings and mixing quartic couplings of the PQ field \cite{Lee:2023dtw,Lee:2024bij}. 

On the other hand, if the PQ symmetry is not broken explicitly,  the angular mode or the axion would be massless before the QCD phase transition, so there exists a nonzero isocurvature perturbation from the angular mode. 
The Planck satellite \cite{planck}  sets a bound on the isocurvature perturbations by
\bea
\beta_{\rm iso} \equiv \frac{P_{\rm iso}(k_*)}{P_{\zeta}(k_*)+P_{\rm iso}(k_*)}<0.038,
\eea
at $95\%$ C.L., with $P_{\zeta}(k_*)=2.1\times 10^{-9}$ at $k_*=0.05\,{\rm Mpc}^{-1}$, leading to
\bea
 \bigg(\frac{\Omega_a}{\Omega_{\rm DM}}\bigg)^2\frac{H^2_I}{\pi^2 \theta^2_* f^2_{a,{\rm eff}}}<8.3\times 10^{-11}.
\eea
The effective decay constant of the axion is large at the horizon exit, namely, $f_{a,{\rm eff}}\simeq 22 M_P$ for $|a|\lesssim 1$ \cite{Lee:2024bij}. Thus, for $\Omega_a=\Omega_{\rm DM}$ and $\theta_*=\pi$, we find that $H_I<1.1\times 10^{15}\,{\rm GeV}$. Therefore, the current bound from the isocurvature perturbation is consistent with our predicted value, $H_I=(2.48\times 10^{14}\,{\rm GeV})\sqrt{r}$ with $r\lesssim 0.0033$ for $|a|\lesssim 1$ in our model.

\section{Gravitational production of Weyl photon dark matter}

In the post-inflationary period of the pole inflation, the coherent oscillation of the inflaton starts, reheating the universe. Assuming that a quartic term in the Einstein frame potential in eq.~(\ref{Epot}) is dominant as in Higgs or PQ pole inflation models, the potential for the canonical inflaton becomes $V_E(\psi)\simeq \lambda_\phi\psi^4$ for which the equation of state for the inflaton during reheating becomes radiation-like \footnote{For the Higgs pole inflation, the detailed analysis on perturbative reheating and some discussion on the perturbation equations were given in Ref.~\cite{Higgspole}, but it would be also important to study the preheating effects in the pole inflation models \cite{Ema:2016dny, Sfakianakis:2018lzf}. But, we postpone a detailed analysis on preheating to a future work.}, i.e. $\omega_\psi=\frac{1}{3}$. Then, in the presence of the decay and scattering of the inflaton, the post-inflationary dynamics in the perturbative regime is governed by the Boltzmann equations \cite{Higgspole},
\bea
{\dot\rho}_\psi +3(1+w_\psi)H\rho_\psi& =&-(1+w_\psi) \Gamma_\psi \rho_\psi, \\
 {\dot\rho}_R +4H\rho_R &=&(1+w_\psi) \Gamma_\psi \rho_\psi,
\eea
with $H^2=\frac{1}{3M^2_P}(\rho_\psi+\rho_R)$. Here, $\rho_\psi, \rho_R$ are the energy densities for the inflaton and the SM radiation bath, and $\Gamma_\psi$ contains the decay and scattering rates for the inflaton. Then, the evolution of the inflaton energy density is approximately given by $\rho_\psi\simeq \rho_{\psi,{\rm end}}(a/a_{\rm end})^{-3(1+w_\psi)}$ and the SM radiation energy density is also approximated to
\begin{widetext}
\bea
\rho_R(a)\simeq \frac{8M_P\Gamma_\psi\sqrt{ \rho_{\psi,{\rm end}}}}{\sqrt{3}(5-3w_\psi)}\bigg(\frac{a}{a_{\rm end}}\bigg)^{-\frac{3}{2}(1+w_\psi)}\bigg(1- \bigg(\frac{a}{a_{\rm end}}\bigg)^{-\frac{1}{2}(5-3w_\psi)}\bigg).
\eea
Thus, the reheating temperature $T_{\rm RH}$ is determined by $\rho_\psi=\rho_R=\frac{\pi^2 g_{\rm RH}}{30} T^4_{\rm RH}$. 
\end{widetext}

\begin{figure}[t]
\centering
\includegraphics[width=0.50\textwidth,clip]{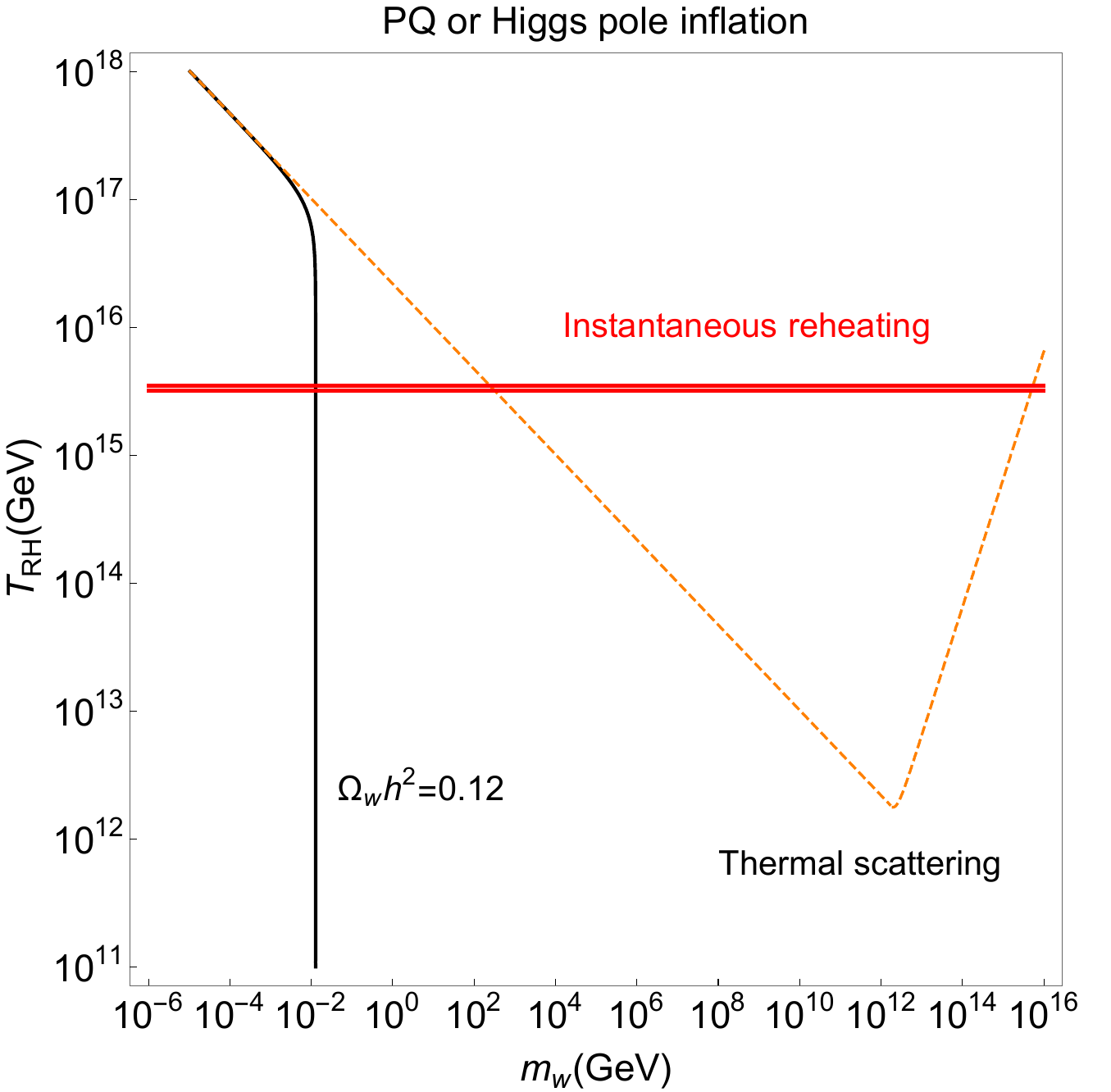} 
\caption{Reheating temperature $T_{\rm RH}$ vs Weyl photon mass $m_w$ in the pole inflation with a quartic potential. The orange dashed line corresponds to the contour satisfying the correct relic density with thermal scattering only, while the black line is the case after both inflaton scattering and thermal scattering are included. A small discrepancy between the cases for PQ and Higgs pole inflation models  is not shown.
} 
\label{Fig:relic}
\end{figure}

On the other hand, the Weyl photon can be produced from the gravitational scattering of the inflaton as well as the gravitational scattering of the radiation during or after reheating. Thus, solving the Boltzmann equations for the number density of the Weyl photon during and after reheating \cite{reheating}, we obtain the relic abundance of the Weyl photon as
\bea
\Omega_w h^2& =&1.6\times 10^8\, m_w \Big(\frac{g_{*,0}}{g_{\rm RH}}\Big) \Big( Y_{w,{\rm inflaton}}  \nonumber \\
&&+Y_{w,{\rm thermal}}+ Y_{w,{\rm reheating}}\Big),
\eea
with
\bea
Y_{w,{\rm inflaton}} &\simeq& \frac{2.53  g_{RH}^{3/4} \lambda^{3/4}_\phi \psi^3_{\rm end}}{M_P^3},  \\
Y_{w,{\rm thermal}} &\simeq& \frac{56469 T^3_{\rm RH}}{128\pi^6\sqrt{10 g_{\rm RH}}M_P^3}, \\
Y_{w,{\rm reheating}} &\simeq & \left\{ \begin{array}{c}Y_{w,{\rm thermal}}, \quad {\rm PQ\,\,inflation},  \vspace{0.2cm} \\  \frac{18759}{18823}\,Y_{w,{\rm thermal}}, \quad {\rm Higgs\,\,inflation}.  \end{array}\right.
\eea
Here, $g_{*,0}=3.91$ and $g_{\rm RH}=106.75$ are taken, and $\psi_{\rm end}$ is the inflation field value at the end of inflation, set to $\psi_{\rm end}\simeq 1.5\,M_P$, and $\lambda_\phi=10^{-11}$ at reheating from the CMB normalization. In the case of the Higgs pole inflation, we kept only the SM fermions and gauge bosons for thermal scattering during reheating ($Y_{w,{\rm reheating}}$) as the Higgs fields have large field-dependent masses. We also note that the contribution from the inflaton scattering is independent of the reheating temperature \cite{G-portal}, unlike the case with a matter-like inflaton during reheating \cite{Yongtang,HiggsR2}. 

In Fig.~\ref{Fig:relic}, the black lines show the contours for the correct relic density for the Weyl photon dark matter in $T_{\rm RH}$ vs $m_w$ for the PQ or Higgs pole inflations. The orange dashed lines correspond to the correct relic density when only the thermal scattering processes during and after reheating are taken into account. The temperature for instantaneous reheating becomes maximal, as shown in the red lines. We find that the contribution from the inflaton scattering is independent of the reheating temperature and dominant as compared to the thermal scattering \cite{PIDM}. The observed dark matter abundance is accounted for by the Weyl photon of about $10\,{\rm MeV}$ mass.  A small difference between the relic abundances in the PQ and Higgs pole inflations is not visible in Fig.~\ref{Fig:relic}.

\section{Conclusions}

We presented the microscopic origin of the pole inflation with scalar fields in the $N$-dimensional multiplet in Weyl gravity.  We showed that the broken $SO(1,N)$ isometry in the field space in combination with the Weyl symmetry restricts the form of the Lagrangian in the Jordan frame such that the vacuum energy is dominant during inflation near the pole of the kinetic term for the inflaton in the Einstein frame. An explicit breaking of the $SO(1,N)$ symmetry to $SO(N)$ is necessary for a slow-roll inflation near the pole.

From the Higgs or PQ inflation models with Weyl symmetry, we found that one parameter family of the pole inflation exists, depending on the overall coefficient $a$ of the Weyl covariant derivatives for scalar fields. The same coefficient not only makes the inflationary predictions for the spectral index and the tensor-to-scalar ratio varying, being compatible with the Planck data, but also determines the mass of the Weyl gauge field.  
As compared to the $T$ $\alpha$-attractor models \cite{linde}, our results show an interesting consequence that the tensor-to-scalar ratio decreases as the Weyl photon mass, $m^2_w=\frac{6a\, g^2_w}{1+a}$, increases, for $a=[0,1]$.

A successful inflation with Higgs or PQ fields is possible while the isocurvature perturbations of the axion can be suppressed sufficiently in the latter case, due to a large effective axion decay constant during inflation. We also showed that the massive Weyl photon can be a dark matter candidate of about $10\,{\rm MeV}$ mass, which is produced dominantly from the inflaton scattering during reheating.

\textit{Acknowledgments}--- 
We are supported in part by Basic Science Research Program through the National
Research Foundation of Korea (NRF) funded by the Ministry of Education, Science and
Technology (NRF-2022R1A2C2003567).
We acknowledge support by Institut Pascal at Universit\'e Paris-Saclay during the Paris-Saclay Astroparticle Symposium 2024.
%


\begin{thebibliography}{1}%
\makeatletter
\providecommand \@ifxundefined [1]{%
 \@ifx{#1\undefined}
}%
\providecommand \@ifnum [1]{%
 \ifnum #1\expandafter \@firstoftwo
 \else \expandafter \@secondoftwo
 \fi
}%
\providecommand \@ifx [1]{%
 \ifx #1\expandafter \@firstoftwo
 \else \expandafter \@secondoftwo
 \fi
}%
\providecommand \natexlab [1]{#1}%
\providecommand \enquote  [1]{``#1''}%
\providecommand \bibnamefont  [1]{#1}%
\providecommand \bibfnamefont [1]{#1}%
\providecommand \citenamefont [1]{#1}%
\providecommand \href@noop [0]{\@secondoftwo}%
\providecommand \href [0]{\begingroup \@sanitize@url \@href}%
\providecommand \@href[1]{\@@startlink{#1}\@@href}%
\providecommand \@@href[1]{\endgroup#1\@@endlink}%
\providecommand \@sanitize@url [0]{\catcode `\\12\catcode `\$12\catcode
  `\&12\catcode `\#12\catcode `\^12\catcode `\_12\catcode `\%12\relax}%
\providecommand \@@startlink[1]{}%
\providecommand \@@endlink[0]{}%
\providecommand \url  [0]{\begingroup\@sanitize@url \@url }%
\providecommand \@url [1]{\endgroup\@href {#1}{\urlprefix }}%
\providecommand \urlprefix  [0]{URL }%
\providecommand \Eprint [0]{\href }%
\providecommand \doibase [0]{http://dx.doi.org/}%
\providecommand \selectlanguage [0]{\@gobble}%
\providecommand \bibinfo  [0]{\@secondoftwo}%
\providecommand \bibfield  [0]{\@secondoftwo}%
\providecommand \translation [1]{[#1]}%
\providecommand \BibitemOpen [0]{}%
\providecommand \bibitemStop [0]{}%
\providecommand \bibitemNoStop [0]{.\EOS\space}%
\providecommand \EOS [0]{\spacefactor3000\relax}%
\providecommand \BibitemShut  [1]{\csname bibitem#1\endcsname}%
\let\auto@bib@innerbib\@empty
\bibitem [{Note1()}]{Note1}%
  \BibitemOpen
  \bibinfo {note} {For the Higgs pole inflation, the detailed analysis on
  perturbative reheating and some discussion on the perturbation equations were
  given in Ref.~\cite {Higgspole}, but it would be also important to study the
  preheating effects in the pole inflation models \cite {Ema:2016dny,
  Sfakianakis:2018lzf}. But, we postpone a detailed analysis on preheating to a
  future work.}\BibitemShut {Stop}%
\end{thebibliography}%


\begin{thebibliography}{999}



\bibitem{Higgsinflation}
F.~L.~Bezrukov and M.~Shaposhnikov,
Phys. Lett. B \textbf{659} (2008), 703-706
doi:10.1016/j.physletb.2007.11.072
[arXiv:0710.3755 [hep-th]].


\bibitem{unitarity}
C.~P.~Burgess, H.~M.~Lee and M.~Trott,
JHEP \textbf{09} (2009), 103
doi:10.1088/1126-6708/2009/09/103
[arXiv:0902.4465 [hep-ph]];
C.~P.~Burgess, H.~M.~Lee and M.~Trott,
JHEP \textbf{07} (2010), 007
doi:10.1007/JHEP07(2010)007
[arXiv:1002.2730 [hep-ph]];
J.~L.~F.~Barbon and J.~R.~Espinosa,
Phys. Rev. D \textbf{79} (2009), 081302
doi:10.1103/PhysRevD.79.081302
[arXiv:0903.0355 [hep-ph]];
M.~P.~Hertzberg,
JHEP \textbf{11} (2010), 023
doi:10.1007/JHEP11(2010)023
[arXiv:1002.2995 [hep-ph]].


\bibitem{sigmamodels}
G.~F.~Giudice and H.~M.~Lee,
Phys. Lett. B \textbf{694} (2011), 294-300
doi:10.1016/j.physletb.2010.10.035
[arXiv:1010.1417 [hep-ph]].



\bibitem{conformal}
Y.~Ema, K.~Mukaida and J.~van de Vis,
JHEP \textbf{11} (2020), 011
doi:10.1007/JHEP11(2020)011
[arXiv:2002.11739 [hep-ph]];
H.~M.~Lee and A.~G.~Menkara,
JHEP \textbf{09} (2021), 018
doi:10.1007/JHEP09(2021)018
[arXiv:2104.10390 [hep-ph]].


\bibitem{linde}
R.~Kallosh and A.~Linde,
JCAP \textbf{12} (2013), 006
doi:10.1088/1475-7516/2013/12/006
[arXiv:1309.2015 [hep-th]].


\bibitem{Kallosh:2013yoa}
R.~Kallosh, A.~Linde and D.~Roest,
JHEP \textbf{11} (2013), 198
doi:10.1007/JHEP11(2013)198
[arXiv:1311.0472 [hep-th]].


\bibitem{Broy:2015qna}
B.~J.~Broy, M.~Galante, D.~Roest and A.~Westphal,
JHEP \textbf{12} (2015), 149
doi:10.1007/JHEP12(2015)149
[arXiv:1507.02277 [hep-th]].


\bibitem{Galante:2014ifa}
M.~Galante, R.~Kallosh, A.~Linde and D.~Roest,
Phys. Rev. Lett. \textbf{114} (2015) no.14, 141302
doi:10.1103/PhysRevLett.114.141302
[arXiv:1412.3797 [hep-th]].


\bibitem{Terada:2016nqg}
T.~Terada,
Phys. Lett. B \textbf{760} (2016), 674-680
doi:10.1016/j.physletb.2016.07.058
[arXiv:1602.07867 [hep-th]].


\bibitem{Weyl0}
L.~Smolin,
Nucl. Phys. B \textbf{160} (1979), 253-268
doi:10.1016/0550-3213(79)90059-2





\bibitem{Weyl1}
D.~M.~Ghilencea and H.~M.~Lee,
Phys. Rev. D \textbf{99} (2019) no.11, 115007
doi:10.1103/PhysRevD.99.115007
[arXiv:1809.09174 [hep-th]].


\bibitem{Weyl2}
Y.~Tang and Y.~L.~Wu,
JCAP \textbf{03} (2020), 067
doi:10.1088/1475-7516/2020/03/067
[arXiv:1912.07610 [hep-ph]].


\bibitem{Weyl3}
S.~Aoki and H.~M.~Lee,
Phys. Rev. D \textbf{108} (2023) no.3, 035045
doi:10.1103/PhysRevD.108.035045
[arXiv:2207.05484 [hep-ph]].



\bibitem{Higgspole}
S.~Clery, H.~M.~Lee and A.~G.~Menkara,
JHEP \textbf{10} (2023), 144
doi:10.1007/JHEP10(2023)144
[arXiv:2306.07767 [hep-ph]].


\bibitem{Lee:2023dtw}
H.~M.~Lee, A.~G.~Menkara, M.~J.~Seong and J.~H.~Song,
JHEP \textbf{05} (2024), 295
doi:10.1007/JHEP05(2024)295
[arXiv:2310.17710 [hep-ph]].


\bibitem{Lee:2024bij}
H.~M.~Lee, A.~G.~Menkara, M.~J.~Seong and J.~H.~Song,
[arXiv:2408.17013 [hep-ph]].





\bibitem{planck}
Y.~Akrami \textit{et al.} [Planck],
Astron. Astrophys. \textbf{641} (2020), A10
doi:10.1051/0004-6361/201833887
[arXiv:1807.06211 [astro-ph.CO]].



\bibitem{keck}
P.~A.~R.~Ade \textit{et al.} [BICEP and Keck],
Phys. Rev. Lett. \textbf{127} (2021) no.15, 151301
doi:10.1103/PhysRevLett.127.151301
[arXiv:2110.00483 [astro-ph.CO]].

\bibitem{Ema:2016dny}
Y.~Ema, R.~Jinno, K.~Mukaida and K.~Nakayama,
JCAP \textbf{02} (2017), 045
doi:10.1088/1475-7516/2017/02/045
[arXiv:1609.05209 [hep-ph]].


\bibitem{Sfakianakis:2018lzf}
E.~I.~Sfakianakis and J.~van de Vis,
Phys. Rev. D \textbf{99} (2019) no.8, 083519
doi:10.1103/PhysRevD.99.083519
[arXiv:1810.01304 [hep-ph]].



\bibitem{reheating}
M.~A.~G.~Garcia, K.~Kaneta, Y.~Mambrini and K.~A.~Olive,
JCAP \textbf{04} (2021), 012
doi:10.1088/1475-7516/2021/04/012
[arXiv:2012.10756 [hep-ph]].


\bibitem{G-portal}
S.~Clery, Y.~Mambrini, K.~A.~Olive and S.~Verner,
Phys. Rev. D \textbf{105} (2022) no.7, 075005
doi:10.1103/PhysRevD.105.075005
[arXiv:2112.15214 [hep-ph]];


\bibitem{Yongtang}
Y.~Tang and Y.~L.~Wu,
Phys. Lett. B \textbf{774} (2017), 676-681
doi:10.1016/j.physletb.2017.10.034
[arXiv:1708.05138 [hep-ph]].


\bibitem{HiggsR2}
S.~Aoki, H.~M.~Lee, A.~G.~Menkara and K.~Yamashita,
JHEP \textbf{05} (2022), 121
doi:10.1007/JHEP05(2022)121
[arXiv:2202.13063 [hep-ph]].


\bibitem{PIDM}
M.~Garny, A.~Palessandro, M.~Sandora and M.~S.~Sloth,
JCAP \textbf{02} (2018), 027
doi:10.1088/1475-7516/2018/02/027
[arXiv:1709.09688 [hep-ph]].






\end{thebibliography}

\newpage
\pagenumbering{gobble} 
\section*{Supplemental Material}
Here we provide the details on the redefined Weyl gauge field in the gauge-fixed Lagrangian, the slow-roll parameters for the pole inflation and the bound on isocurvature perturbations for the PQ inflation.

{\it Redefinition of the Weyl gauge field}:
First, we consider the part of the Lagrangian for the Weyl gauge field as
\bea
{\cal L}_w &=& -\frac{1}{4} w_{\mu\nu}w^{\mu\nu}+\frac{1}{2} (m^2_w-a g^2_w \phi^2_i) w_\mu w^\mu \nonumber \\
&&+\frac{1}{2}a g_w w_\mu \partial^\mu\phi^2_i, 
\eea
which can be rewritten as
\bea
{\cal L}_w &=&  -\frac{1}{4} {\widetilde w}_{\mu\nu}{\widetilde w}^{\mu\nu} +\frac{1}{2}  (m^2_w - a g^2_w \phi_i^2) {\widetilde w}_\mu  {\widetilde w}^\mu \nonumber \\
&&-\frac{1}{8} a^2 g^2_w\cdot  \frac{(\partial_\mu \phi^2_i)^2}{ (m^2_w - a g^2_w \phi_i^2) }, \label{Lagw}
\eea
thanks to the field definition with
\bea
  {\widetilde w}_\mu \equiv  w_\mu-\frac{1}{2g_w} \partial_\mu \ln   (m^2_w - a g^2_w \phi_i^2).
\eea

{\it Slow-roll parameters for the pole inflation}:
The slow-roll parameters from the inflaton potential for the Higgs pole inflation, in eq.~(\ref{Vinf}), are given by
\bea
\epsilon &=&\frac{1}{2}  \bigg(\frac{V'_E}{V_E}\bigg)^2 \nonumber \\
&=&\frac{32}{\langle\chi^2\rangle} \bigg[  \sinh\Big(\frac{2\psi}{\langle\chi\rangle}\Big)\bigg]^{-2} ,  \label{ep} \\
\eta &=&  \frac{V^{\prime\prime}_E}{V_E} \nonumber \\
&=&-\frac{16}{\langle\chi^2\rangle}\bigg[  \cosh\Big(\frac{2\psi}{\langle\chi\rangle}\Big)-4 \bigg]  \bigg[\sinh\Big(\frac{2\psi}{\langle\chi\rangle}\Big)\bigg]^{-2}. \label{eta}
\eea
The number of efoldings is 
\bea
N&=&\int^{\psi_*}_{\psi_e} \frac{ {\rm sgn} (V'_E)d\psi}{\sqrt{2\epsilon}} \nonumber \\
&=&\frac{\langle\chi^2\rangle}{16}\bigg[ \cosh\Big(\frac{2\psi_*}{\langle\chi\rangle}\Big)- \cosh\Big(\frac{2\psi_e}{\langle\chi\rangle}\Big)  \bigg]. \label{efold}
\eea
Here, $\psi_*, \psi_e$ are the values of the Higgs boson at horizon exit and the end of inflation, respectively. Here, we note that $\epsilon=1$ determines $\phi_e$.
As a result, using  the above slow-rill parameters and $N\simeq \frac{\langle\chi^2\rangle}{16}\,  \cosh\Big(\frac{2\psi_*}{\langle\chi\rangle}\Big)$ for $\psi_*\gg \langle\chi\rangle $ during inflation, we obtain the slow-roll parameters at horizon exit in terms of the number of efoldings as
\bea
\epsilon_* &\simeq & \frac{32 \langle \chi^2\rangle}{256N^2-\langle\chi^4\rangle},  \\
\eta_* &\simeq& -\frac{64(4N-\langle\chi^2\rangle)}{256N^2-\langle\chi^4\rangle}.
\eea
Thus, we get the spectral index in terms of the number of efoldings, as follows, 
\bea
n_s&=&1-6\epsilon_*+2\eta_* \nonumber \\
&=& 1-\frac{64(8N+\langle\chi^2\rangle)}{256N^2-\langle\chi^4\rangle}. \label{sindex}
\eea
Moreover, the tensor-to-scalar ratio at horizon exit is
\bea
r=16\epsilon_* =  \frac{512 \langle\chi^2\rangle}{256N^2-\langle\chi^4\rangle}. \label{ratio}
\eea

{\it Bounds on the isocurvature perturbations}:
The power spectrum of the isocurvature perturbation at the horizon exit of the mode $k_*$ during inflation depends on the axion abundance, $\Omega_a$, is given by
\bea
P_{\rm iso}(k_*)&=&\bigg(\frac{1}{\Omega_{\rm DM}} \frac{\partial \Omega_a}{\partial\theta_*}\bigg)^2 \langle\delta \theta_*^2\rangle \nonumber \\
&=& \bigg(\frac{\Omega_a}{\Omega_{\rm DM}}\bigg)^2\frac{4}{\theta^2_*}\langle\delta \theta_*^2\rangle
\eea
where $\Omega_{\rm DM}$ is the dark matter abundance at present and the fluctuation of the initial angle is given by
\bea
 \langle\delta \theta_*^2\rangle= \frac{1}{f^2_{a,{\rm eff}}} \,\bigg(\frac{H_I}{2\pi}\bigg)^2.
\eea
with $f_{a,{\rm eff}}\equiv \sqrt{6} M_P\big|\sinh\big(\frac{\phi_*}{\langle\chi\rangle}\big)\big|$.
The bound on the isocurvature perturbations from Planck satellite \cite{planck} is given by
\bea
\beta_{\rm iso} \equiv \frac{P_{\rm iso}(k_*)}{P_{\zeta}(k_*)+P_{\rm iso}(k_*)}<0.038,
\eea
at $95\%$ C.L., with $P_{\zeta}(k_*)=2.1\times 10^{-9}$ at $k_*=0.05\,{\rm Mpc}^{-1}$, namely,
\bea
 \bigg(\frac{\Omega_a}{\Omega_{\rm DM}}\bigg)^2\frac{H^2_I}{\pi^2 \theta^2_* f^2_{a,{\rm eff}}}<8.3\times 10^{-11}.
\eea

\vspace{0.2in}

\end{document}